\documentclass[prb,amsmath,amsfonts,amssymb,reprint,unsortedaddress,floatfix]{revtex4-1}
\usepackage{dcolumn}
\usepackage{graphicx}
\usepackage{bm}
\usepackage{color}

\begin{document}
\title{Seniority-based coupled cluster theory}
\author{Thomas M. Henderson}
\affiliation{Department of Chemistry, Rice University, Houston, TX 77005-1892}
\affiliation{Department of Physics and Astronomy, Rice University, Houston, TX 77005-1892}

\author{Ireneusz W. Bulik}
\affiliation{Department of Chemistry, Rice University, Houston, TX 77005-1892}

\author{Tamar Stein}
\affiliation{Department of Chemistry, Rice University, Houston, TX 77005-1892}

\author{Gustavo E. Scuseria}
\affiliation{Department of Chemistry, Rice University, Houston, TX 77005-1892}
\affiliation{Department of Physics and Astronomy, Rice University, Houston, TX 77005-1892}
\date{\today}

\begin{abstract}
Doubly occupied configuration interaction (DOCI) with optimized orbitals often accurately describes strong correlations while working in a Hilbert space much smaller than that needed for full configuration interaction.  However, the scaling of such calculations remains combinatorial with system size.  Pair coupled cluster doubles (pCCD) is very successful in reproducing DOCI energetically, but can do so with low polynomial scaling ($N^3$, disregarding the two-electron integral transformation from atomic to molecular orbitals).  We show here several examples illustrating the success of pCCD in reproducing both the DOCI energy and wave function, and show how this success frequently comes about.  What DOCI and pCCD lack are an effective treatment of dynamic correlations, which we here add by including higher-seniority cluster amplitudes which are excluded from pCCD.  This frozen pair coupled cluster approach is comparable in cost to traditional closed-shell coupled cluster methods with results that are competitive for weakly correlated systems and often superior for the description of strongly correlated systems.
\end{abstract}
\maketitle

\section{Introduction}
The coupled cluster (CC) family of methods\cite{Paldus1999,Bartlett2007,BartlettShavitt} offer a powerful wave function approach to the description of weakly correlated systems, to the point that the accurate treatment of such systems is essentially routine: provided that the system is not too large, one can simply apply coupled cluster with single and double excitations\cite{CCSD} (CCSD) or CCSD plus perturbative triple excitations,\cite{CCSDT} which we refer to as CCSD(T).  The same, unfortunately, cannot be said for the coupled cluster treatment of strongly correlated systems, for which traditional single-reference methods such as CCSD or CCSD(T) may fail badly.  Much progress has been made in multi-reference coupled cluster theory,\cite{Bartlett2007} to be sure, but the techniques are by no means black box or computationally inexpensive.  Continued developments of coupled cluster techniques for strongly correlated systems is essential.

In 2013, Ayers and coworkers made a surprising disocvery along these lines: a method which they refer to as the antisymmetric product of 1-reference orbital geminals\cite{Limacher2013,Limacher2014,Tecmer2014,Boguslawski2014} (AP1roG) and which we will refer to as pair coupled cluster doubles\cite{Stein2014} (pCCD) provides a remarkably reasonable description of the strong correlations for a wide variety of systems.  What makes this so surprising is that pCCD looks like coupled cluster doubles (CCD) restricted to include only those excitations which preserve electron pairs, but pCCD, unlike CCD, seems to be able to describe strong correlations.  Why should a simplification of a fundamentally single-reference method be able to describe multi-reference problems?

In this manuscript, we seek to do three things.  First, we want to provide a self-contained description of pCCD, with all the equations one needs to implement the approach.  Second, we wish to offer some perspective on the method's successes.  Third, we wish to go beyond pCCD and include some of the dynamic correlations which pCCD does not provide.  To accomplish this, however, we first must discuss doubly occupied configuration interaction and orbital seniority.

\section{Seniority and Doubly Occupied Configuration Interaction}
Pair coupled cluster theory is based on the concept of the seniority of a determinant.  The seniority is the number of unpaired electrons.  The idea is simple: every spinorbital $\phi_p$ is paired with one and only one other spinorbital, $\phi_{\bar{p}}$, and the seniority of a determinant is the number of spinorbital pairs which between them contain only one electron.  Loosely speaking, seniority is related to the number of broken electron pairs.

In this work, as in our previous work on the subject,\cite{Stein2014} we restrict ourselves to singlet pairing, in which the orbitals that are paired are the two spinorbitals corresponding to the same spatial orbital.  In that case, the seniority operator is just
\begin{equation}
\Omega = N - 2 \, D
\end{equation}
where $N$ is the number operator
\begin{equation}
N = \sum_p \left(c_{p_\uparrow}^\dagger \, c_{p_\uparrow} + c_{p_\downarrow}^\dagger \, c_{p_\downarrow}\right) = \sum_p \left(n_{p_\uparrow} + n_{p_\downarrow}\right)
\end{equation}
and $D$ is a double-occupancy operator
\begin{equation}
D = \sum_p c_{p_\uparrow}^\dagger \, c_{p_\downarrow}^\dagger \, c_{p_\downarrow} \, c_{p_\uparrow} = \sum_p \, n_{p_\uparrow} \, n_{p_\downarrow}.
\end{equation}
Throughout this work, we will use indices $i$, $j$, $k$, $l$ for occupied spatial orbitals, $a$, $b$, $c$, $d$ for virtual spatial orbitals, and $p$, $q$, $r$, $s$ for general spatial orbitals.

It is important to notice that seniority depends on which orbitals we use to define the double-occupancy operator $D$, because a unitary transformation which mixes the orbitals leaves $N$ invariant but changes the form of $D$.  If we define seniority with respect to the molecular orbitals of the restricted Hartree-Fock (RHF) determinant $|\textrm{RHF}\rangle$, then we see that the RHF determinant is a seniority eigenfunction and has seniority zero.  If we define seniority with respect to a different basis, this need not be true.  It is also important to note that seniority is not a symmetry of the molecular Hamiltonian -- $[H,\Omega] \neq 0$ -- which means that the exact wave function is not an eigenfunction of $\Omega$.

The utility of the seniority concept comes from using it as an alternative to organize Hilbert space.\cite{Bytautas2011}  Conventionally, we describe determinants in terms of their excitation level, which we can extract from the particle-hole number operator
\begin{equation}
2 \, N_{ph} = \sum_a \, \left(n_{a_\uparrow} + n_{a_\downarrow}\right) + \sum_i \, \left(2 - n_{i_\uparrow} - n_{i_\downarrow}\right).
\end{equation}
As with seniority, the excitation level is neither orbitally invariant (because defining particles and holes with respect to a different Fermi vacuum changes the excitation level) nor a symmetry of the Hamiltonian, but it nevertheless provides a valuable framework within which we can organize Hilbert space and solve the Schr\"odinger equation in a subspace.  The exact wave function is generally a linear combination of determinants of all possible excitation levels, and similarly it is generally a linear combination of determinants of all possible seniorities.  The success of single-reference coupled cluster theory for weakly correlated systems is grounded on the fact that the coupled cluster expansion in terms of particle-hole excitations out of the Hartree-Fock determinant converges rapidly toward full configuration interaction (FCI).  The ground state of weakly correlated systems, then, is characterized by having a low number of particle-holes.

\begin{figure}[t]
\includegraphics[width=0.45\textwidth]{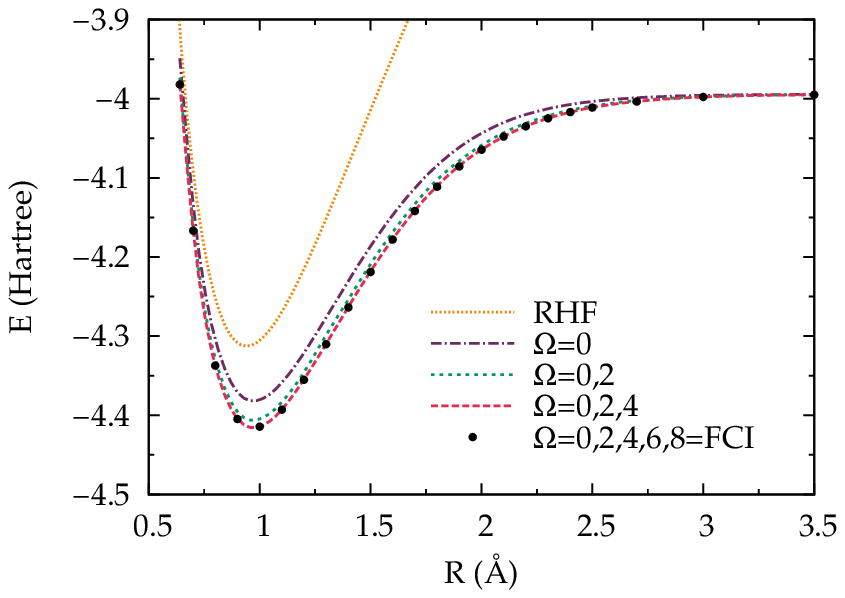}
\\
\includegraphics[width=0.45\textwidth]{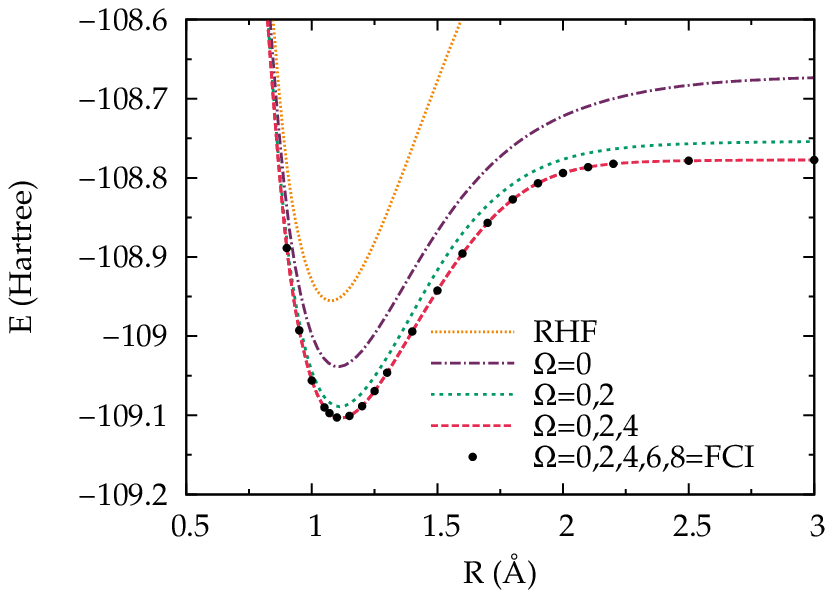}
\caption{Top panel: Dissociation of the equally-spaced H$_8$ chain.  Bottom panel: Dissociation of N$_2$.  Both calculations are done in the cc-pvdz basis set and restrict the CI problem to a minimal active space.  We emphasize that curves are obtained with an RHF wave function.  Results taken from Ref. \onlinecite{Bytautas2011}.
\label{fig:Laimis}}
\end{figure}

We posit that the ground state of strongly correlated systems is characterized by having a \textit{low seniority number in a suitable one-electron basis}.  One can test this by defining configuration interaction (CI) restricted to the zero seniority sector of Hilbert space, which we refer to as doubly occupied configuration interaction (DOCI).\cite{Allen1962,Smith1965,Weinhold1967,Couty1997,Kollmar2003,Bytautas2011}   Because DOCI is not invariant to the orbitals with respect to which seniority is defined, we optimize this choice energetically.  This is analagous to optimizing the identity of the reference determinant in an excitation-truncated CI calculation, or with optimizing the orbitals in CAS-SCF, though DOCI is generally size consistent.  As we and others have shown, DOCI with orbital optimization provides a valuable tool for the description of strong correlations.  This can be shown in Fig. \ref{fig:Laimis}, which shows that DOCI gives the correct limit in the dissociation of the equally spaced H$_8$ chain and gives most of the strong correlation in N$_2$ as well.  Note that these plots are generated using a minimal active space to remove, to the degree possible, dynamic correlation at dissociation.

The chief drawback of DOCI is that of computational cost: the number of determinants with $\Omega=0$ is just the square root of the number of all determinants with a given particle number, so the cost of DOCI is the square root of the cost of full CI.  Worse yet, it is more difficult to use symmetry to eliminate determinants from DOCI than it is to eliminate determinants from FCI.  For example, every DOCI determinant is a spin singlet with our singlet pairing scheme, so we cannot use spin symmetry to reduce the number of determinants to be included.  In practice, DOCI calculations on systems with more than a few dozen electrons are prohibitively expensive.

This is where pCCD enters the picture: pCCD generally provides results which for the molecular Hamiltonian are nearly indistinguishable from those of DOCI, but whereas the computational cost of DOCI scales combinatorially with system size, the cost of pCCD scales as $\mathcal{O}(N^3)$.

\section{Pair Coupled Cluster Doubles}
In pCCD, we write the wave function as
\begin{equation}
|\Psi\rangle = \mathrm{e}^T |0\rangle
\end{equation}
where $|0\rangle$ is a closed-shell reference determinant and
\begin{equation}
T = \sum_{ia} t_i^a \, P_a^\dagger \, P_i
\end{equation}
in terms of the pair operators $P_a^\dagger$ and $P_i$, where generically
\begin{equation}
P_q^\dagger = c_{q_\uparrow}^\dagger \, c_{q_\downarrow}^\dagger
\end{equation}
with the singlet pairing we are using.  As usual, one can insert this ansatz into the Schr\"odinger equation to get
\begin{subequations}
\begin{align}
E &= \langle 0| \bar{H} | 0\rangle,
\label{Eqns:pCCD_energy}
\\
0 &= \langle 0| P_i^\dagger \, P_a \, \bar{H} |0 \rangle,
\label{Eqns:pCCD_amplitudes}
\end{align}
\label{Eqns:pCCD}
\end{subequations}
where the similarity transformed Hamiltonian $\bar{H}$ is given by
\begin{equation}
\bar{H} = \mathrm{e}^{-T} \, H \, \mathrm{e}^T.
\end{equation}
In AP1roG, one instead writes
\begin{subequations}
\begin{align}
E &= \langle 0| H \, \mathrm{e}^T | 0\rangle,
\label{Eqns:AP1roG_energy}
\\
E \, \langle 0| P_i^\dagger \, P_a \, \mathrm{e}^T |0\rangle &= \langle 0| P_i^\dagger \, P_a \, H \, \mathrm{e}^T |0 \rangle,
\label{Eqns:AP1roG_amplitudes}
\end{align}
\label{Eqns:AP1roG}
\end{subequations}
but because
\begin{subequations}
\begin{align}
\langle 0| \mathrm{e}^{-T}  &= \langle 0|
\\
\langle 0 | P_i^\dagger \, P_a \, \mathrm{e}^{-T} &= \langle 0 | P_i^\dagger \, P_a - t_i^a \, \langle 0|
\\
 &= \langle 0 | P_i^\dagger \, P_a - \langle 0| P_i^\dagger \, P_a \, \mathrm{e}^T |0\rangle \, \langle 0|
\nonumber
\end{align}
\end{subequations}
one can see that Eqns. \ref{Eqns:pCCD_energy} and \ref{Eqns:AP1roG_energy} are identical, and consequently so too are Eqns. \ref{Eqns:pCCD_amplitudes} and \ref{Eqns:AP1roG_amplitudes}.

Explicitly, the pCCD energy and amplitudes are given by
\begin{subequations}
\begin{align}
E &= \langle 0| H |0\rangle + \sum_{ia} t_i^a \, v^{ii}_{aa}
\\
0 &= v_{ii}^{aa} + 2 \, \Big(f^a_a - f^i_i- \sum_j v^{jj}_{aa} \, t_j^a - \sum_b v^{ii}_{bb} \, t_i^b\Big) t_i^a
\label{Eqn:TEqns}
\\
  &- 2 \, \Big(2 \, v^{ia}_{ia} - v^{ia}_{ai} - v^{ii}_{aa} \, t_i^a \Big) t_i^a
\nonumber
\\
 &+ \sum_b v^{aa}_{bb} \, t_i^b
  + \sum_j v_{ii}^{jj} \, t_j^a
  + \sum_{jb} v^{jj}_{bb} \, t_j^a \, t_i^b 
\nonumber
\end{align}
\end{subequations}
where $f^p_q$ is an element of the Fock operator and $v^{pq}_{rs} = \langle \phi_p \, \phi_q | V_{ee} | \phi_r \, \phi_s\rangle$ is a two-electron integral in Dirac notation.  As promised, these equations can be solved in $\mathcal{O}(N^3)$ computational cost with the aid of  the intermediate $y_i^j = \sum_b v^{jj}_{bb} \, t_i^b$.

As with traditional CC methods, we can define a left-hand eigenvector $\langle \mathcal{L}|$ of $\bar{H}$ in CI-like fashion:
\begin{equation}
\langle \mathcal{L} | = \langle 0| (1+Z)
\end{equation}
where
\begin{equation}
Z = \sum_{ia} z^i_a \, P_i^\dagger \, P_a.
\end{equation}
Then the expectation value of $\bar{H}$ is
\begin{equation}
\mathcal{E} = \langle 0| (1+Z) \, \bar{H} | 0\rangle = \langle 0| (1+Z) \, \mathrm{e}^{-T} \, H \, \mathrm{e}^T |0 \rangle.
\end{equation}
The equations for the amplitudes $t_i^a$ are just
\begin{equation}
0 = \frac{\partial \mathcal{E}}{\partial z^a_i}
\end{equation}
and guarantee by their satisfaction that
\begin{equation}
\mathcal{E} = \langle 0| \bar{H} |0\rangle 
\end{equation}
for any value of $Z$; similarly, we obtain the amplitudes $z^a_i$ from
\begin{equation}
0 = \frac{\partial \mathcal{E}}{\partial t_i^a}.
\end{equation}
We find that the $z$ equations are
\begin{align}
0 &= v^{ii}_{aa}
   + 2 \, \Big(f^a_a - f^i_i - \sum_j v^{jj}_{aa} \, t_j^a - \sum_b v^{ii}_{bb} \, t_i^b\Big) \, z^i_a
\\
  &- 2 \, \Big(2 \, v^{ia}_{ia} - v^{ia}_{ai} - 2 \, v^{ii}_{aa} \, t_i^a\Big) \, z^i_a
\nonumber
\\
  &- 2 \, v^{ii}_{aa} \, \Big(\sum_j z^j_a \, t_j^a + \sum_b z^i_b \, t_i^b\Big)
\nonumber
\\
  &+ \sum_b v^{bb}_{aa} \, z^i_b
   + \sum_j v_{jj}^{ii} \, z^j_a
   + \sum_{jb} t_j^b \, \left(v^{ii}_{bb} \, z^j_a + v^{jj}_{aa} \, z^i_b\right).
\nonumber
\end{align}
Again, these can be solved in $\mathcal{O}(N^3)$ time.  We should emphasize that the pCCD energy and amplitude equations for both $T$ and $Z$ can be extracted from the usual RHF-based CCD\cite{Scuseria1987b,Scuseria1988} by simply retaining only the pair amplitudes $t_{ii}^{aa}$ and $z^{aa}_{ii}$ which we have here written as simply $t_i^a$ and $z^a_i$ for compactness of notation and to emphasize that the pCCD $t$ and $z$ amplitudes are two-index quantities.  In practice, one usually finds that $Z \sim T^\dagger$, as we might expect.  We note in passing that one can readily identify the various channels\cite{Scuseria2008,Scuseria2013} of the CCD amplitude equations in Eqn. \ref{Eqn:TEqns}, where the ladder terms are found on the third line, the ring and crossed-ring terms appear on the second line, and what we have termed the Brueckner or mosaic terms appear on the first line.  For pCCD, the various ring terms decouple, though our limited numerical experience suggests that a pair ring CCD model is not useful.

Like DOCI, pCCD is not invariant to the choice of which orbitals are used to define the pair operators $P_p^\dagger$.  Additionally, pCCD depends on the choice of reference determinant $|0\rangle$.  In order to have a well-defined method, we must provide a way of fixing these choices.  This can be accomplished by orbital optimization,\cite{Scuseria1987,Bozkaya2011} for which purpose we introduce the one-body antihermitian operator
\begin{equation}
\kappa = \sum_{p>q} \sum_\sigma \kappa_{pq} \, \left(c_{p_\sigma}^\dagger \, c_{q_\sigma} - c_{q_\sigma}^\dagger \, c_{p_\sigma}\right)
\end{equation}
which, when exponentiated, creates unitary orbital rotations; here, $\sigma$ indexes spins (\textit{i.e.} $\sigma = \uparrow,\downarrow$).  Note that in contrast to the typical coupled-cluster orbital optimization which includes only occupied-virtual mixing, we must allow \textit{all} orbitals to mix.  We have taken $\kappa$ to be real.

Given the rotation operator, we can simply generalize the energy to
\begin{equation}
\mathcal{E}(\kappa) = \langle 0 | (1+Z) \, \mathrm{e}^{-T} \, \mathrm{e}^{-\kappa} \, H \, \mathrm{e}^{\kappa} \, \mathrm{e}^T |0\rangle
\end{equation}
and make it stationary with respect to $\kappa$, which gives us
\begin{align}
0 &= \left.\frac{\partial \mathcal{E}(\kappa)}{\partial \kappa_{pq}}\right|_{\kappa=0}
\\
 &= \sum_\sigma \langle 0| (1+Z) \, \mathrm{e}^{-T} \, [H, c_{p_\sigma}^\dagger \, c_{q_\sigma} - c_{q_\sigma}^\dagger \, c_{p_\sigma}] \, \mathrm{e}^T |0\rangle
\nonumber
\end{align}
where we work at $\kappa = 0$ by transforming the basis in which we express the Hamiltonian (\textit{i.e.} by transforming the one- and two-electron integrals).  The commutator can be evaluated readily:
\begin{align}
[H, c_{p_\sigma}^\dagger \, c_{q_\sigma}]
 &= \sum_r h^r_p \, c_{r_\sigma}^\dagger \, c_{q_\sigma}
  - \sum_r h^q_r \, c_{p_\sigma}^\dagger \, c_{r_\sigma}
\\
 &+ \sum_{rst} \, \sum_{\sigma^\prime} v^{rs}_{pt} \, c_{r_\sigma}^\dagger \, c_{s_{\sigma^\prime}}^\dagger \, c_{t_{\sigma^\prime}} \, c_{q_\sigma}
\nonumber
\\
 &- \sum_{rst} \, \sum_{\sigma^\prime} v^{qt}_{rs} \, c_{p_\sigma}^\dagger \, c_{t_{\sigma^\prime}}^\dagger \, c_{s_{\sigma^\prime}} \, c_{r_\sigma}
\nonumber
\end{align}
where the Hamiltonian is
\begin{equation}
H = \sum_{pq} \sum_\sigma h^p_q \, c_{p_\sigma}^\dagger \, c_{q_\sigma} + \frac{1}{2} \, \sum_{pqrs} \sum_{\sigma \sigma^\prime} v^{pq}_{rs} \, c_{p_\sigma}^\dagger \, c_{q_{\sigma^\prime}}^\dagger \, c_{s_{\sigma^\prime}} \, c_{r_\sigma}
\end{equation}
in terms of one-electron integrals $h^p_q$ and the two-electron integrals $v^{pq}_{rs}$ previously defined.  The energy gradient is then
\begin{align}
\left.\frac{\partial \mathcal{E}(\kappa)}{\partial \kappa_{pq}}\right|_{\kappa=0}
 &= \Big[\sum_r \left(h^r_p \, \gamma^q_r - h^q_r \, \gamma^r_p\right)
\label{Eqn:OrbGradient}
\\
 &\hspace{2ex}
 + \sum_{rst} \left(v^{rs}_{pt} \, \Gamma^{qt}_{rs} - v^{qt}_{rs} \, \Gamma^{rs}_{pt}\right)\Big] - \left(p \leftrightarrow q\right)
\nonumber
\end{align}
where $\gamma^p_q$ and $\Gamma^{pq}_{rs}$ are one-body and two-body density matrices, given by
\begin{subequations}
\begin{align}
\gamma^p_q &= \sum_\sigma \langle 0| (1+Z) \, \mathrm{e}^{-T} \, c_{q_\sigma}^\dagger \, c_{p_\sigma} \, \mathrm{e}^T |0\rangle,
\\
\Gamma^{pq}_{rs} &= \sum_{\sigma \sigma^\prime} \langle 0| (1+Z) \, \mathrm{e}^{-T} \, c_{r_\sigma}^\dagger \, c_{s_{\sigma^\prime}}^\dagger \, c_{q_{\sigma^\prime}} \, c_{p_\sigma} \, \mathrm{e}^T |0\rangle.
\end{align}
\end{subequations}
We use a Newton-Raphson scheme to minimize the norm of the orbital gradient, which finds an orbital stationary point.  Having found such a point, we check the eigenvalues of the coupled cluster orbital Hessian and, if there is a negative eigenvalue, follow the instability until we find a local energy minimum or saddle point (\textit{i.e.} we look for points with zero gradient and non-negative Hessian).  The analytic formulae for the density matrices and the orbital Hessian are presented in the appendix.  As has been previously pointed out, there are multiple solutions to the orbital optimization equations, and because the optimized orbitals are generally local in character if the system is strongly correlated,\cite{Bytautas2011,Limacher2014,Stein2014} it proves convenient to start from the RHF determinant with localized molecular orbitals.  We should also point out that convergence of the pair amplitude and response equations is greatly aided by using DIIS.\cite{Scuseria1986}  Our Newton-Raphson procedure typically uses the diagonal Hessian and turns on the full analytic Hessian only near convergence; this avoids getting trapped in high energy local minima.

It should be noted here that the one-body density matrix $\bm{\gamma}$ is diagonal in the basis in which we define the pairing.  In other words, the molecular orbitals defining the pCCD $T$ and $Z$ operators are also the natural orbitals of pCCD.  The two-body density matrix $\bm{\Gamma}$ is also very sparse and has a kind of semi-diagonal form where only $\Gamma_{pp}^{qq}$, $\Gamma_{pq}^{pq}$, and $\Gamma_{pq}^{qp}$ are non-zero.  These properties are true both for pCCD and for DOCI (and indeed for any zero-seniority wave function method).  Detailed expressions for the density matrices can be found in the Appendix.

\section{Pair Coupled Cluster and Doubly Occupied Configuration Interaction}
Now that we have given ample detail about pCCD and have introduced DOCI, it will prove useful to compare results from the two methods for a variety of small systems for which the DOCI calculations are feasible.  We will compare the energies from the two approaches, and also look at overlaps of the pCCD and DOCI wave functions; explicitly, we will compute
\begin{equation}
\Delta E = E_\mathrm{pCCD} - E_\mathrm{DOCI}
\label{Eqn:defDE}
\end{equation}
to assess the quality of the pCCD energy and
\begin{equation}
S = \langle 0 | (1+Z) \, \mathrm{e}^{-T} | \mathrm{DOCI}\rangle \, \langle \mathrm{DOCI} | \mathrm{e}^T |0\rangle
\label{Eqn:defS}
\end{equation}
to assess the quality of the pCCD wave functions.  Note that $S \approx 1$ when pCCD is close to DOCI; more explicitly, we have
\begin{equation}
\langle 0 | (1+Z) \, \mathrm{e}^{-T} \, \mathrm{e}^T |0\rangle =1,
\end{equation}
and inserting the projector $|\mathrm{DOCI}\rangle \, \langle \mathrm{DOCI}|$ should not substantially change this value when pCCD and DOCI roughly coincide.  Because pCCD is biorthogonal, we do \textit{not} have $S < 1$; indeed, we will frequently see that $S$ is slightly larger than one.  We emphasize here that both pCCD and DOCI can be symmetry adapted despite having individual orbitals which are not symmetry eigenfunctions, due to the orbital optimization; indeed, for the examples discussed below pCCD with optimized orbitals appears to respect point-group symmetry, though we have found model Hamiltonians for which this is not the case.  We will always compare DOCI and pCCD with the same orbital set (usually orbitals optimized for pCCD).  Spot checks show that typically orbitals optimized for DOCI are virtually indistinguishable from orbitals optimized for pCCD.

All DOCI and pCCD calculations in this section and indeed throughout the manuscript use in-house programs, as do the frozen-pair coupled cluster calculations discussed in Sec. \ref{sec:fpCC}; other calculations used the \textit{Gaussian} program package.\cite{Gaussian}  Throughout, we will use Dunning's cc-pVDZ basis set,\cite{Dunning1989} because we need a sufficiently small basis that the DOCI is computationally tractable, though we will use Cartesian rather than spherical $d$-functions.

We start by noting that for H$_2$, as for any two-electron singlet, pCCD with orbital optimization is exact (and is equivalent to DOCI).  This is just because one can use occupied-virtual rotations to make single excitations in CCSD vanish (in other words, one can do Brueckner coupled cluster doubles) and then pick a virtual-virtual rotation to eliminate the seniority two excitation amplitudes.  One can see this by noting that for a two-electron singlet, we have
\begin{equation}
T = \frac{1}{2} \, \sum_{ab} t_{1,1}^{ab} c_{a_\uparrow}^\dagger \, c_{b_\downarrow}^\dagger \, c_{1_\downarrow} \, c_{1_\uparrow};
\end{equation}
the combination of fermionic antisymmetry and spin symmetry means that $t_{1,1}^{ab} = t_{1,1}^{ba}$, so we can define a real symmetric matrix $M_{ab} = t_{1,1}^{ab}$ which can be diagonalized by a virtual-virtual rotation so that $T$ takes the pCCD form.  Numerically, we find that with optimized orbitals, $E_\mathrm{pCCD} = E_\mathrm{DOCI} = E_\mathrm{FCI}$ and $S = 1$, as we should.

\begin{figure}[t]
\includegraphics[width=0.48\textwidth]{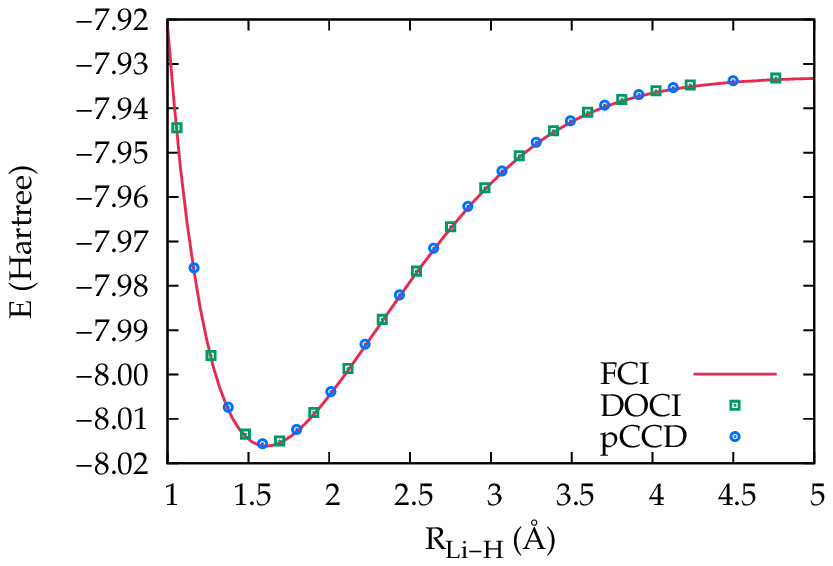}
\\
\includegraphics[width=0.48\textwidth]{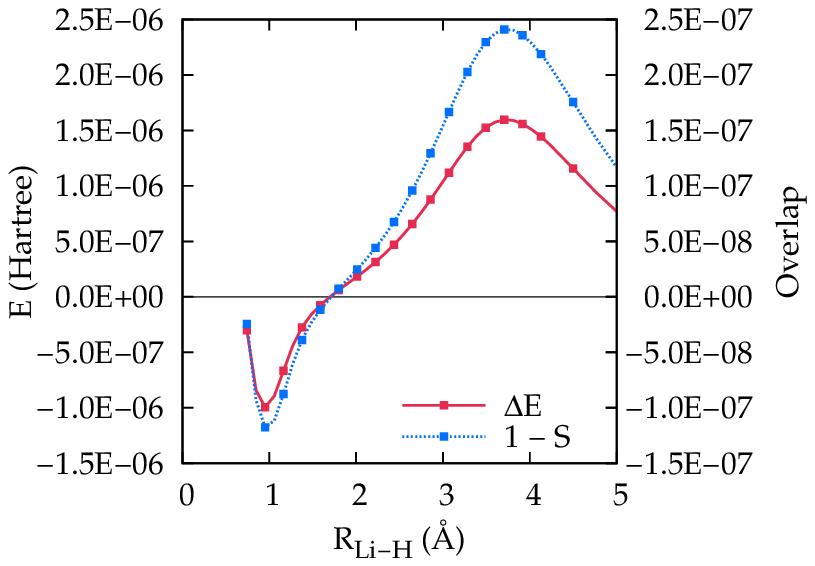}
\caption{Dissociation of LiH.  Top panel: Dissociation energies from FCI, DOCI, and pCCD.  Bottom panel: Difference between DOCI and pCCD energies ($\Delta E$, defined in Eqn. \ref{Eqn:defDE} and measured on the left axis) and in the overlap ($1-S$, measured on the right axis with $S$ defined in Eqn. \ref{Eqn:defS}).
\label{Fig:LiH}}
\end{figure}

In Fig. \ref{Fig:LiH} we show results for the dissociation of LiH.  Because LiH is a quasi-two--electron problem, we would expect DOCI and pCCD to be very accurate in this case.  Indeed, Fig. \ref{Fig:LiH} shows that pCCD and DOCI are energetically indistinguishable and both are essentially superimposable with FCI (errors are on the order of 0.4 mE$_\mathrm{H}$ throughout the dissociation).  Moreover, the DOCI and pCCD wave functions have near unit overlap throughout the dissociation.  This is exactly what we would expect for such a problem.  

\begin{figure}[t]
\includegraphics[width=0.48\textwidth]{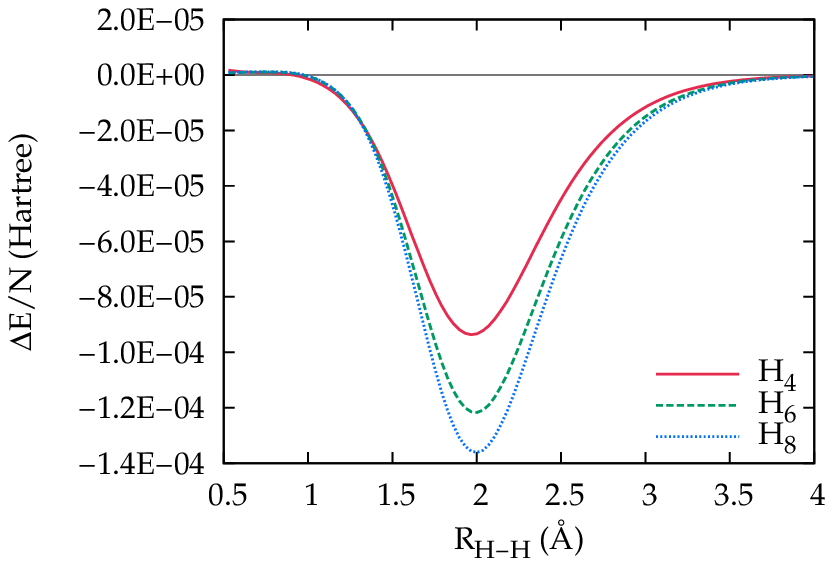}
\\
\includegraphics[width=0.48\textwidth]{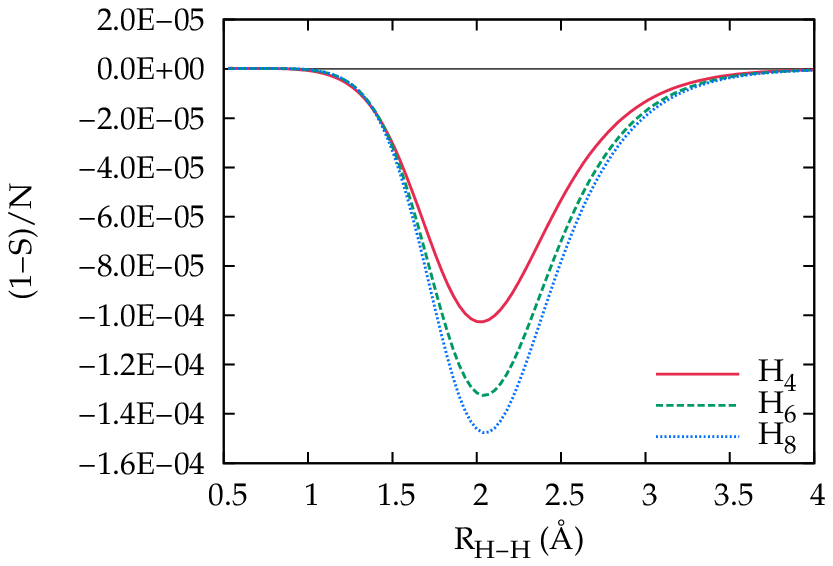}
\caption{Dissociation of equally spaced hydrogen chains.  Top panel: Differences between DOCI and pCCD energies ($\Delta E$, defined in Eqn. \ref{Eqn:defDE}) per electron pair.  Bottom panel: deviations in the overlap ($1-S$, with $S$ defined in Eqn. \ref{Eqn:defS}) per electron pair.
\label{Fig:HChain}}
\end{figure}

We next turn our attention to the dissociation of equally spaced hydrogen chains.  These serve as important prototypes of strongly correlated systems and map in a loose sense to the Hubbard Hamiltonian.\cite{Hubbard1963}  The top panel of Fig. \ref{Fig:HChain} shows the difference between the DOCI and pCCD energies per electron pair, while the bottom panel shows the deviation of the overlap $S$ from unity, again per electron pair.  These results appear to be saturating, though unfortunately the DOCI calculations on H$_{10}$ are impracticably expensive with our code.

\begin{figure}[t]
\includegraphics[width=0.48\textwidth]{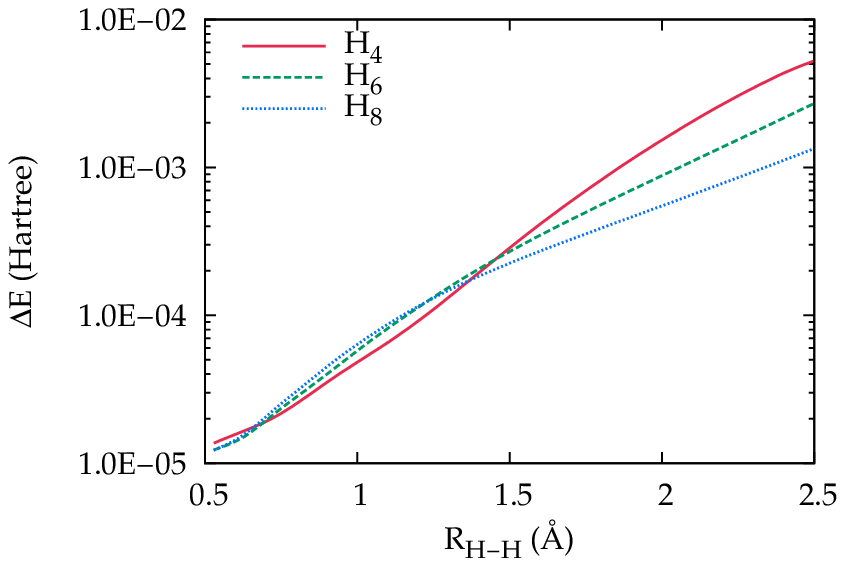}
\\
\includegraphics[width=0.48\textwidth]{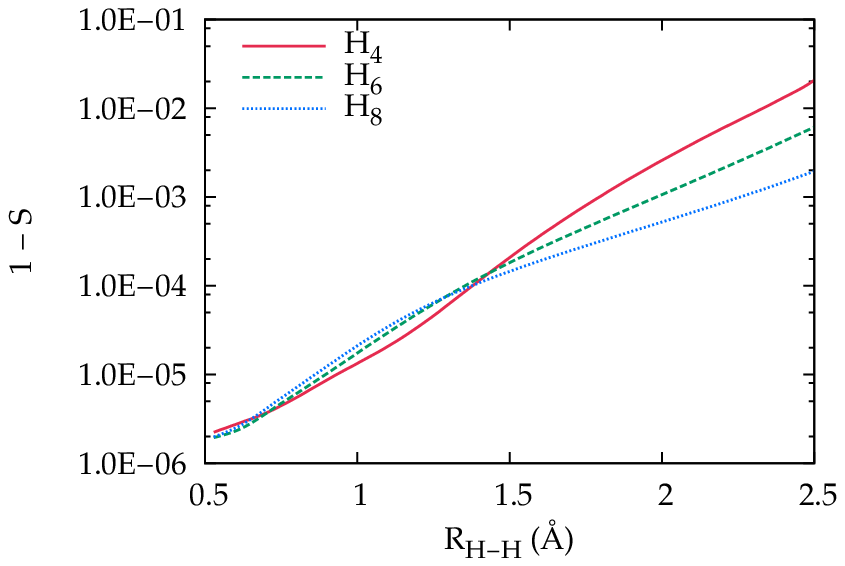}
\caption{Dissociation of equally spaced hydrogen chains in the canonical RHF basis rather than the pCCD-optimized basis used elsewhere.  Top panel: Differences between DOCI and pCCD energies ($\Delta E$, defined in Eqn. \ref{Eqn:defDE}).  Bottom panel: deviations in the overlap ($1-S$, with $S$ defined in Eqn. \ref{Eqn:defS}).
\label{Fig:HChainCan}}
\end{figure}

We should note that while the equivalence between DOCI and pCCD has been established for energetically optimized orbitals, we see the same general behavior when DOCI and pCCD pair canonical RHF orbitals instead, though not to the same degree.  That is, even pairing canonical RHF orbitals rather than optimized orbitals, pCCD and DOCI give energies that agree to within a few milliHartree, with the agreement predictably degrading as the systems become more strongly correlated.  We can see this in hydrogen chains in Fig. \ref{Fig:HChainCan}.  Strangely, the agreement between DOCI and pCCD appears to improve as we move from H$_4$ to H$_6$ to H$_8$ when using canonical RHF orbitals, while in the optimized orbital case we see the opposite behavior.  We should emphasize that the deviations in the energy and overlap in Fig. \ref{Fig:HChainCan} are not shown per electron pair.

\begin{figure}
\includegraphics[width=0.48\textwidth]{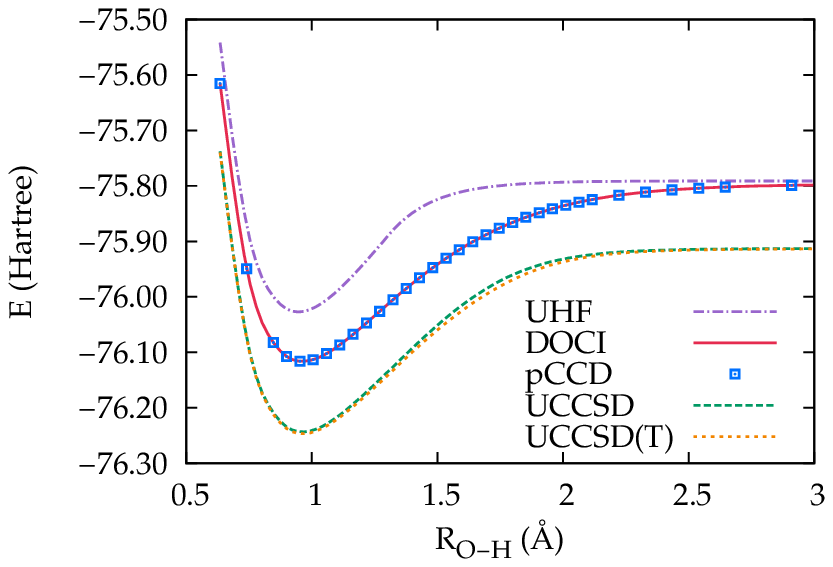}
\\
\includegraphics[width=0.48\textwidth]{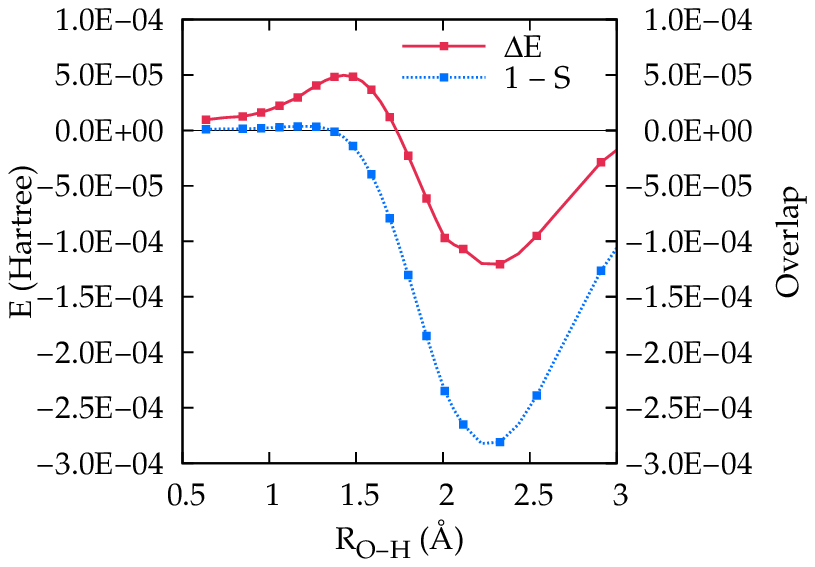}
\caption{Symmetric double dissociation of H$_2$O.  Top panel: Dissociation energies from DOCI and pCCD, as well as from unrestricted Hartree-Fock (UHF) and CCSD and CCSD(T) based thereon.  Bottom panel: Errors in the energy ($\Delta E$, defined in Eqn. \ref{Eqn:defDE} and measured on the left axis) and in the overlap ($1-S$, measured on the right axis with $S$ defined in Eqn. \ref{Eqn:defS}).
\label{Fig:H2O}}
\end{figure}

Our next example is the symmetric double dissociation of H$_2$O, as shown in Fig. \ref{Fig:H2O}.  Again, pCCD and DOCI provide nearly identical energies throughout the dissociation process, and the overlaps of the pCCD and DOCI wave functions are large.  The coincidence of DOCI and pCCD, in other words, is true not just for one pair of strongly correlated electrons, but for two pairs as well.  At dissociation, DOCI and pCCD give essentially the unrestricted Hartree-Fock (UHF) result, despite being closed-shell wave functions, though as we shall see later, this is somewhat fortuitous.  These methods miss a significant amount of the correlation compared to UHF-based CCSD and CCSD(T); the dynamic correlation, then, is clearly not well described.

\begin{figure}
\includegraphics[width=0.48\textwidth]{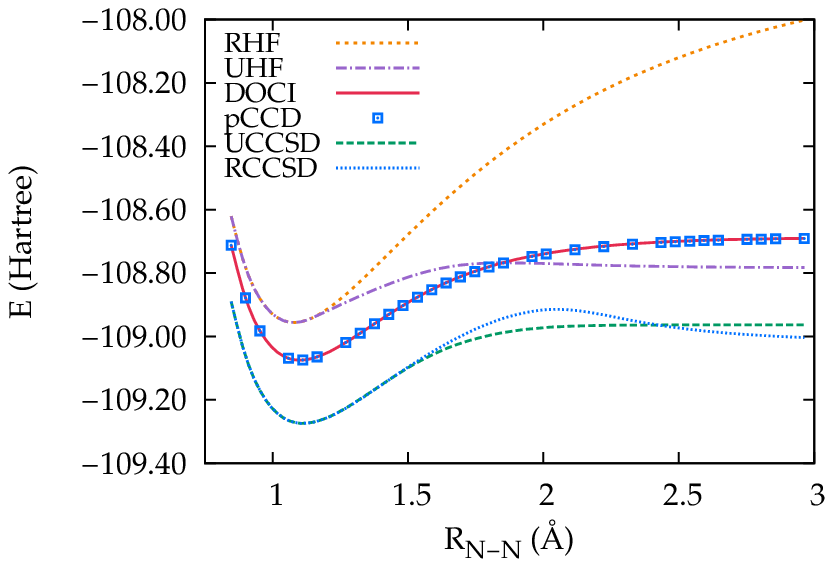}
\\
\includegraphics[width=0.48\textwidth]{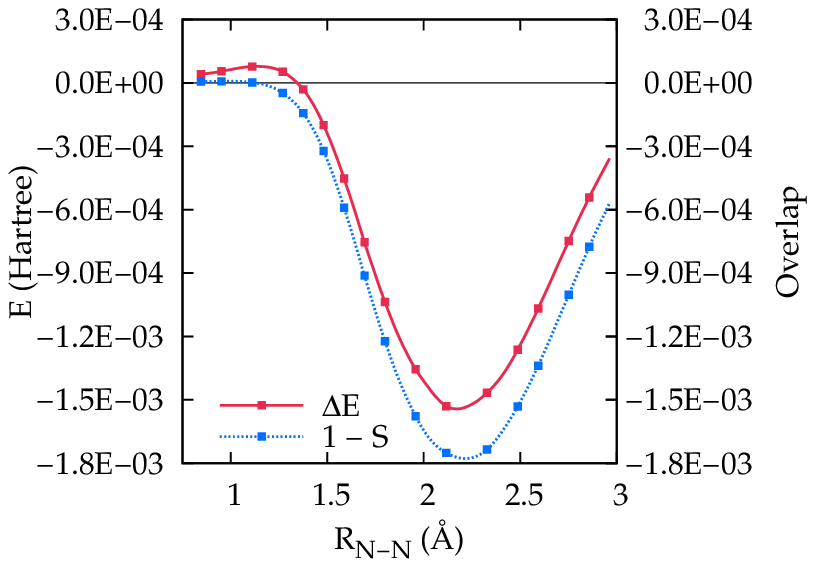}
\caption{Dissociation of N$_2$.  Top panel: Dissociation energies from DOCI and pCCD, as well as from RHF, UHF, and RHF- and UHF-based CCSD. Bottom panel: Errors in the energy ($\Delta E$, defined in Eqn. \ref{Eqn:defDE} and measured on the left axis) and in the overlap ($1-S$, measured on the right axis with $S$ defined in Eqn. \ref{Eqn:defS}).
\label{Fig:N2}}
\end{figure}

Similar conclusions can be reached from examining the dissociation of N$_2$.  As Fig. \ref{fig:Laimis} reveals, DOCI does not give all the strong correlation needed to dissociate the triple bond in N$_2$ correctly, but does offer substantial improvements over RHF.  We see similar results in Fig. \ref{Fig:N2}.  In these calculations, we froze the nitrogen $1s$ core orbitals after the orbital optimization and compare the frozen-core DOCI to the frozen-core pCCD.  We also note that our procedure of repeatedly following instabilities in the pCCD orbital Hessian led to an unphysical reference determinant for which the pCCD broke down; we have thus used a stationary point rather than a minimum of the pCCD energy functional to define the reference.  Our results reiterate that pCCD and DOCI get most but not all of the strong correlation in N$_2$, and fail to account for the dynamic correlation effectively.  Nonetheless, even for this triple bond we see that DOCI and pCCD have close agreement.

\begin{table}
\caption{Energies and overlaps in the neon atom.  Here, $E_\mathrm{Ref}$ denotes the energy of the reference determinant.  We show results for both the optimized determinant for pCCD and for the canonical RHF determinant as a reference.
\label{tab:Neon}}
\begin{tabular}{lcc}
\hline\hline
            &  Optimized     &  Canonical \\
\hline
$E_\mathrm{Ref}$   &  -128.488 823  & -128.488 866  \\
$E_\mathrm{DOCI}$  &  -128.559 677  & -128.546 705  \\
$E_\mathrm{pCCD}$  &  -128.559 674  & -128.546 701  \\
$E_\mathrm{CCSD}$  &  -128.683 931  & -128.683 958  \\
$1 - S$     &  $1.43 \times 10^{-7}$  
            &  $1.16 \times 10^{-7}$  \\
\hline\hline
\end{tabular}
\end{table}

One can see that DOCI and pCCD do not describe dynamic correlation particularly well by considering the neon atom, as seen in Tab. \ref{tab:Neon}.  While DOCI and pCCD are in excellent agreement with one another, they only retrieve about 36\% of the correlation energy even after orbital optimization, with optimized orbitals very close to the canonical RHF molecular orbitals.  The bulk of the correlations must then involve determinants of higher seniority.  In order to remedy this deficiency, we turn to what we call frozen-pair coupled cluster,\cite{Stein2014} as we will describe shortly.

First, however, it may be instructive to take a closer look at the $T$-amplitudes of pCCD and the CI coefficients of DOCI, to understand why the two methods coincide so neatly.  Often, what we find, as in the examples above, is that the pCCD $T$-amplitudes are such that each occupied orbital is strongly correlated with at most one virtual orbital, so that each row of the matrix $t_i^a$ has at most one large entry, while most of the amplitudes are small.  The DOCI vector follows this same basic structure, which is unsurprising since the DOCI and pCCD wave functions are essentially the same.  In these cases, pCCD and DOCI are similar to a kind of perfect pairing wave function.\cite{Hurley1953,Goddard1978,VanVoorhis2002}  For example, for the stretched H$_2$O case, the pCCD and DOCI wave functions are qualitatively
\begin{equation}
|\Psi\rangle \approx |\mathrm{O}_{1s}^2 \, \mathrm{O}_\mathrm{lp}^4 \, \left(\mathrm{OH}_\sigma^2 - \alpha \, \mathrm{OH}_{\sigma^\star}^2\right)^2 \rangle
\end{equation}
where $\alpha$ approaches 1 at dissociation and where $\mathrm{O}_{1s}$, $\mathrm{O}_\mathrm{lp}$, $\mathrm{OH}_\sigma$, and $\mathrm{OH}_{\sigma^\star}$ respectively denote the oxygen $1s$ orbital, oxygen lone-pair orbitals, OH bonding orbitals, and OH antibonding orbitals.  In the case of stretched H$_2$O, it is the small deviations from this perfect pairing structure which cause the energy to be close to the UHF limit.  That is, the only wave function amplitudes larger than $\sim 0.05$ correspond to excitations from an OH bonding orbital into its antibonding orbital, but correlating the bonding orbitals alone yields an energy somewhat above the sum of restricted open-shell Hartree-Fock atomic energies.  Thus, we might not expect pCCD to describe strong correlations beyond those accessible with the perfect pairing structure, even though we must emphasize that the pCCD wave function is not inherently limited to this form.

Indeed, it is important to note that we have found cases in the repulsive Hubbard Hamiltonian\cite{Hubbard1963} for which neither pCCD nor DOCI adopt a perfect pairing structure, yet the two methods still agree closely.  We also note that for the attractive pairing Hamiltonian\cite{Henderson2014} or the attractive Hubbard Hamiltonian (results not shown), one can find instances in which pCCD does not resemble DOCI.  In these cases, the DOCI coefficients and the pCCD amplitudes are dense and neither DOCI nor pCCD displays a perfect pairing structure.  While pCCD and DOCI include a perfect pairing wave function as a special case, they are more general methods.  The fact that pCCD closely resembles DOCI seems a key feature of fermionic repulsive Hamiltonians like the molecular one.

\section{Frozen Pair Coupled Cluster
\label{sec:fpCC}}
The basic idea of frozen pair coupled cluster is very simple.  One could imagine decomposing the $T_2$ double-excitation operator into a pair part $T_2^{(0)}$ and a non-pair part $\tilde{T}_2$; one would then solve the pCCD equations for the pair amplitudes and then solve the usual CCD equations without allowing the pair amplitudes to change.  Note that the non-pair operator $\tilde{T}_2$ creates seniority non-zero determinants, which we rely upon to provide the dynamic correlation which pCCD lacks; $\tilde{T}_2$ on a seniority zero determinant returns a linear combination of determinants with seniorities two and four.  Note also that the Fock operator for orbital-optimized pCCD is in general neither diagonal nor in the semicanonical form which diagonalizes the occupied-occupied and virtual-virtual blocks, so the full non-canonical form of the amplitude equations must be used.  This is not a concern for pCCD, where only the diagonal elements of the (generally non-diagonal) Fock operator contribute to the amplitude equations.

What we have described above we would call frozen pair CCD (fpCCD).  One could of course extend this basic idea to include single excitations and triple or higher excitations in the cluster operator.  What we wish to do here is to briefly consider frozen pair coupled cluster with single-, double-, and triple-excitation amplitudes (fpCCSDT).\cite{Noga1987,Scuseria1988b}

\begin{figure}
\includegraphics[width=0.5\textwidth]{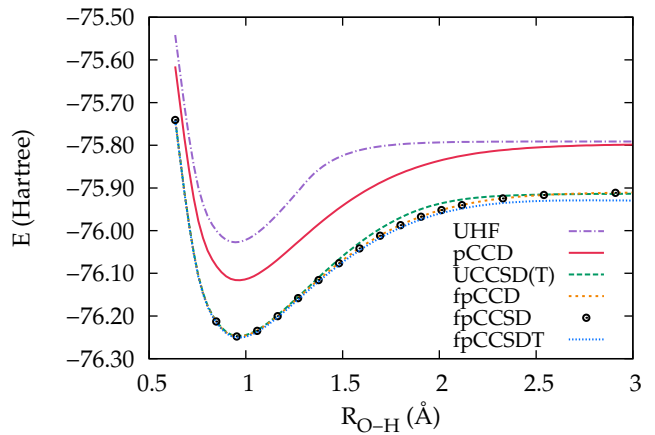}
\caption{Frozen pair symmetric double dissociation of H$_2$O.
\label{Fig:fpH2O}}
\end{figure}

In Fig. \ref{Fig:fpH2O}, we show the symmetric double dissociation of H$_2$O, this time with the frozen pair approximation.  The effect of single excitations is in this case small (fpCCD and fpCCSD give similar results) and fpCCSD gives results fairly similar to the UHF-based CCSD and CCSD(T) curves.  Adding full triple excitations in fpCCSDT gives larger correlation at dissociation and probably overcorrelates somewhat.

\begin{figure}
\includegraphics[width=0.5\textwidth]{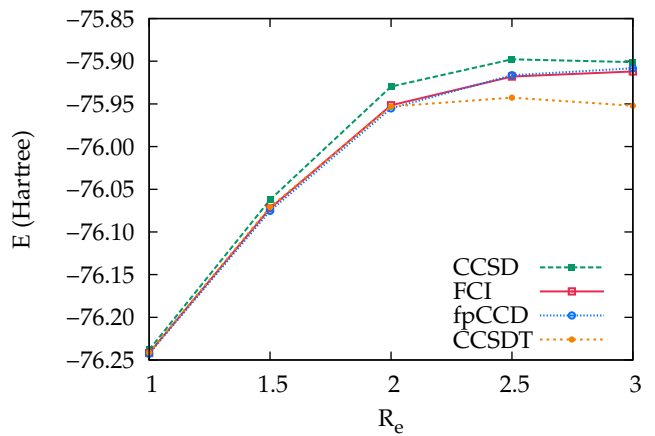}
\caption{Symmetric double dissociation of H$_2$O at 110$^\circ$ bond angle, with frozen pair coupled cluster and traditional coupled cluster methods.  FCI and CCSD data taken from Ref. \onlinecite{Olsen1996}.  All results use closed-shell (restricted) wave functions.
\label{Fig:H2OWeird}}
\end{figure}

For comparison purposes, we show results from FCI and RHF-based CCSD and CCSDT in Fig. \ref{Fig:H2OWeird}.  These calculations fix the H-O-H bond angle at 110$^{\circ}$ rather than at the 104.474$^{\circ}$ used in our other calculations, and use spherical $d$ functions; the CCSD, CCSDT, and FCI data are taken from Ref. \onlinecite{Olsen1996}.  We see that as one stretches the bond, CCSD and CCSDT go through a maximum and turn over; for larger bond lengths, we would expect CCSD and CCSDT to overcorrelate more.  In contrast, fpCCD is coincidentally very close to FCI, and while fpCCSD and fpCCSDT overcorrelate somewhat more, they provide sensibly-shaped dissociation curves without requiring symmetry breaking.

\begin{table}
\caption{Energies in the neon atom.  Here, $E_\mathrm{Ref}$ denotes the energy of the reference determinant.
\label{tab:Neon2}}
\begin{tabular}{lc}
\hline\hline
Method              &  Energy        \\
\hline
$E_\mathrm{Ref}$      &  -128.488 823  \\
$E_\mathrm{pCCD}$     &  -128.559 674  \\
$E_\mathrm{fpCCD}$    &  -128.687 585  \\
$E_\mathrm{CCD}$      &  -128.683 851  \\
$E_\mathrm{fpCCSD}$   &  -128.687 619  \\
$E_\mathrm{CCSD}$     &  -128.683 931  \\
$E_\mathrm{fpCCSDT}$  &  -128.688 497  \\
$E_\mathrm{CCSDT}$    &  -128.685 089  \\
\hline\hline
\end{tabular}
\end{table}

Table \ref{tab:Neon2} shows fpCCD and fpCCSD results for the neon atom.  While pCCD undercorrelates significantly compared to CCSD, making the frozen pair approximation yields results that differ from those without freezing $T_2^{(0)}$ by about 4 milliHartree.  As with the double dissociation of H$_2$O, frozen pair coupled cluster overcorrelates slightly.

\begin{figure}
\includegraphics[width=0.5\textwidth]{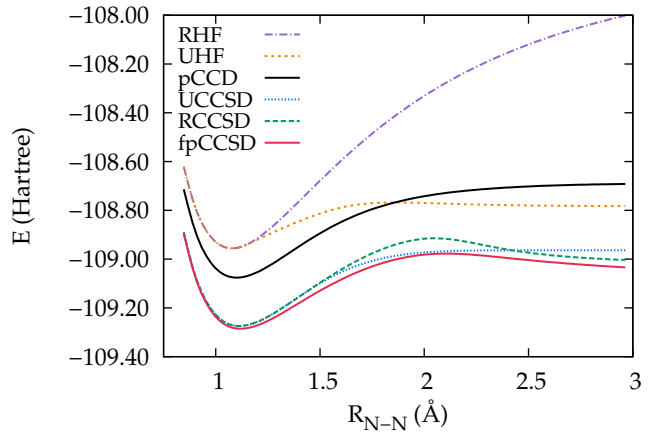}
\caption{Dissociation of N$_2$ with various coupled cluster methods.
\label{Fig:fpN2}}
\end{figure}

As a final example, we consider fpCCSD for the dissociation of N$_2$, as seen in Fig. \ref{Fig:fpN2}.  As should by now be familiar, fpCCSD gives a reasonable accounting for dynamic correlation but overcorrelates somewhat.  Both fpCCSD and RHF-based CCSD break down for large bond lengths, and have an artificial bump in the dissociation curve; while fpCCSD does not eliminate this unphysical effect, it at least mitigates it somewhat.

Our results show that frozen pair coupled cluster should be understood as an easy way to incorporate the reasonable pCCD description of strong correlation while retaining much of the ability of traditional coupled cluster to also describe dynamic correlation.  However, while easy to implement and conceptually simple, it is also important to note that a frozen pair full coupled cluster approach would give the wrong answer.  In other words, in the exact theory one must clearly allow the zero-seniority $T_2$ amplitudes to relax from their pCCD values.  In practice, fpCCSD should allow for a reasonable description of both strongly and weakly correlated systems at essentially the cost of a CCSD calculation, without breaking spin symmetry, although fpCCSD would be expected to break down somewhat for cases where pCCD is unable to capture all the strong correlations, as is the case with N$_2$.  For two-electron singlets, fpCCD and fpCCSD are both exact, because as we have previously noted, pCCD is already FCI, which implies that $T_1$ and the non-zero seniority parts of $T_2$ vanish.

\section{Conclusions}
While traditional coupled cluster theory is highly successful for the description of weakly correlated systems, it generally fails to describe strong correlation.  Paradoxically, by simply eliminating the vast bulk of the cluster operator, one can form pair coupled cluster doubles, which accurately reproduces DOCI, and to the extent that DOCI can describe strong correlations, so too can pCCD.  Moreover, pCCD accomplishes this task with mean-field computational scaling for the coupled cluster part.  Not only does pCCD reproduce the DOCI energy, it also reproduces the DOCI wave function.  The DOCI wave function, in other words, is essentially factorizable into the pCCD form.  Loosely, this can be accomplished because, upon orbital optimization, the pCCD and DOCI wave functions studied in this work adopt a perfect--pairing-like structure.

While pCCD can describe strong correlations, it is much less successful at modeling dynamic correlation, which apparently requires the breaking of electron pairs to obtain higher seniority determinants when we define pairs in terms of the spatial orbitals in a particle-hole representation.  Using pCCD to obtain the zero-seniority part of the cluster operator and then solving the traditional coupled cluster equations for the rest of the amplitudes yields frozen-pair coupled cluster, which seems to be able to describe both weakly and strongly correlated systems with reasonable accuracy and with a computational cost not much different from that of standard coupled cluster methods.

Of course pCCD is not a panacea and there are occasions when pCCD fails to account for the strong correlation present in the DOCI wave function, although we have not seen such a case for the molecular Hamiltonian.  Likewise, it is possible that the DOCI form is too restricted to allow for a complete description of the strong correlations present, as appears to happen in the dissociation of N$_2$, for example.  In such cases, the frozen-pair coupled-cluster approach would be of less utility.  We speculate that it may be possible to include these strong correlations by generalizing the pairing structure to non-singlet pairing, so that the pairs included in pCCD and DOCI are not just the two electrons in the same spatial orbital.  Regardless, we hope that pCCD and its frozen pair extensions will be useful tools for the description of both weakly and strongly correlated systems without the need for symmetry breaking or higher excitation operators.

\begin{acknowledgments}
This work was supported by the National Science Foundation (CHE-1110884).  GES is a Welch Foundation chair (C-0036).  T.S. is an awardee of the Weizmann Institute of Science -- National Postdoctoral Award Program for Advancing Women in Science.  We would like to thank Carlos Jim\'enez-Hoyos for helpful discussion.
\end{acknowledgments}

\appendix
\section{Density Matrices and Orbital Hessian}
For completeness, we include here expressions for the pCCD density matrices and orbital rotation Hessian; together with the orbital rotation gradient of Eqn. \ref{Eqn:OrbGradient}, these provide everything needed for the Newton-Raphson algorithm we use for orbital optimization.

Recall that the energy is written as
\begin{equation}
\mathcal{E}(\kappa) = \langle 0 | (1+Z) \, \mathrm{e}^{-T} \, \mathrm{e}^{-\kappa} \, H \, \mathrm{e}^{\kappa} \, \mathrm{e}^T |0\rangle
\end{equation}
with
\begin{equation}
\kappa = \sum_{p>q} \sum_\sigma \kappa_{pq} \, \left(c_{p_\sigma}^\dagger \, c_{q_\sigma} - c_{q_\sigma}^\dagger \, c_{p_\sigma}\right)
\end{equation}
where the orbital rotation is given by the unitary transformation $\exp(\kappa)$.  At every step of the Newton-Raphson scheme, we solve for $\kappa$, build $\exp(\kappa)$ which rotates to a new orbital basis, transform the integrals, and begin a new iteration.  

We have already seen that the gradient is simply
\begin{equation}
\left.\frac{\partial \mathcal{E}(\kappa)}{\partial \kappa_{pq}}\right|_{\kappa=0}
  = \mathcal{P}_{pq} \, \sum_\sigma \langle [H, c_{p_\sigma}^\dagger \, c_{q_\sigma}] \rangle
\end{equation}
where $\mathcal{P}_{pq}$ is a permutation operator $\mathcal{P}_{pq} = 1 - \left( p\leftrightarrow q\right)$ and the notation for the expectation value means
\begin{equation}
\langle \mathcal{O} \rangle = \langle 0| (1+Z) \, \mathrm{e}^{-T} \, \mathcal{O} \, \mathrm{e}^T |0\rangle.
\end{equation}
Similarly, the Hessian is
\begin{align}
H_{pq,rs}
 &= \left.\frac{\partial^2 \mathcal{E}(\kappa)\hfill}{\partial \kappa_{pq} \, \partial \kappa_{rs}}\right|_{\kappa=0}
\\
 &= \frac{1}{2} \, \mathcal{P}_{pq} \, \mathcal{P}_{rs} \, \sum_{\sigma,\eta} \langle [[H, c_{p_\sigma}^\dagger \, c_{q_\sigma}], c_{r_\eta}^\dagger \, c_{s_\eta}] \rangle
\nonumber
\\
 &+ \frac{1}{2} \, \mathcal{P}_{pq} \, \mathcal{P}_{rs} \, \sum_{\sigma,\eta} \langle [[H, c_{r_\eta}^\dagger \, c_{s_\eta}], c_{p_\sigma}^\dagger \, c_{q_\sigma}] \rangle
\nonumber
\end{align}
where $\eta$ is another spin index.  We obtain
\begin{widetext}
\begin{align}
H_{pq,rs}
 = \mathcal{P}_{pq} \, \mathcal{P}_{rs} \, 
 &  \Big\{\frac{1}{2} \, \sum_u 
      \left[\delta_{qr} \, \left(h^u_p \, \gamma^s_u + h^s_u \, \gamma^u_p\right)
          + \delta_{ps} \, \left(h^u_r \, \gamma^q_u + h^q_u \, \gamma^u_r\right)
      \right]
    - \left(h^s_p \, \gamma^q_r + h^q_r \, \gamma^s_p\right)
\\
 &+ \frac{1}{2} \, \sum_{tuv}
      \left[\delta_{qr} \, \left(v^{uv}_{pt} \, \Gamma^{st}_{uv} + v^{st}_{uv} \, \Gamma^{uv}_{pt}\right)
          + \delta_{ps} \, \left(v^{qt}_{uv} \, \Gamma^{uv}_{rt} + v^{uv}_{rt} \, \Gamma^{qt}_{uv}\right)
      \right]
\nonumber
\\
 &+ \sum_{uv} \left(v^{uv}_{pr} \, \Gamma^{qs}_{uv} + v^{qs}_{uv} \, \Gamma^{uv}_{pr}\right)
  - \sum_{tu} \left(v^{st}_{pu} \, \Gamma^{qu}_{rt} + v^{ts}_{pu} \, \Gamma^{qu}_{tr} + v^{qu}_{rt} \, \Gamma^{st}_{pu} + v^{qu}_{tr} \, \Gamma^{ts}_{pu}\right)
\Big\}.
\nonumber
\end{align}
\end{widetext}

The one-particle density matrix we have defined as
\begin{equation}
\gamma^p_q = \sum_\sigma \langle 0| (1+Z) \, \mathrm{e}^{-T} \, c_{q_\sigma}^\dagger \, c_{p_\sigma} \, \mathrm{e}^T |0\rangle.
\end{equation}
Because $T$ and $Z$ both preserve the seniority of the wave function, and the reference $|0\rangle$ has seniority zero, it is immediately clear that the one-particle density matrix is diagonal in the basis in which we have defined the pairing; the optimized orbital basis for pCCD, in other words, is also its natural orbital basis.  We then have
\begin{subequations}
\begin{align}
\gamma^j_i &= 2 \, \left(1 - x^j_i\right) \, \delta_{ij},
\\
\gamma^b_a &= 2 \, x^b_a \, \delta_{ab},
\\
\gamma^i_a &= \gamma^a_i = 0,
\end{align}
\end{subequations}
where $\delta_{pq}$ is the Kronecker delta and where we have defined
\begin{subequations}
\begin{align}
x_i^j &= \sum_a t_i^a \, z^j_a,
\\
x_a^b &= \sum_i t_i^b \, z^i_a.
\end{align}
\end{subequations}
Recall that $i$ and $a$ are respectively occupied and virtual orbital indices.

Similar considerations show that the two-particle density matrix is also sparse in the natural orbital basis.  The non-zero elements of the two-particle density matrix are
\begin{equation}
\Gamma_{ii}^{jj} = 2 \, \left[x_i^j + \delta_{ij} \, \left(1 - 2 \, x_i^i\right)\right],
\end{equation}
\begin{equation}
\Gamma_{ii}^{aa} = 2 \, \left[t_i^a + x_i^a - 2 \, t_i^a \, \left(x_a^a + x_i^i - t_i^a \, z^i_a\right)\right],
\end{equation}
\begin{equation}
\Gamma_{aa}^{ii} = 2 \, z^i_a,
\end{equation}
\begin{equation}
\Gamma_{aa}^{bb} = 2 \, x_a^b,
\end{equation}
\begin{equation}
\Gamma_{ij}^{ij} = 4 \, \left(1 - x_i^i - x_j^j\right) + 2 \, \delta_{ij} \, \left(3 \, x_i^i - 1\right),
\end{equation}
\begin{equation}
\Gamma_{ia}^{ia} = \Gamma_{ai}^{ai} = 4 \, \left(x_a^a - t_i^a \, z^i_a\right),
\end{equation}
\begin{equation}
\Gamma_{ab}^{ab} = 2 \, \delta_{ab} \, x_a^a,                 
\end{equation}
\begin{equation}
\Gamma_{pq}^{qp} \underset{q \neq p}{=} - \frac{1}{2} \, \Gamma_{pq}^{pq}.
\end{equation}
We have defined the additional intermediate
\begin{equation}
x_i^a = \sum_{jb} t_i^b \, t_j^a \, z^j_b.
\end{equation}

Note that the sparsity of the one- and two-particle density matrices allows one to considerably reduce the cost of evaluating the Hessian.

\bibliography{pCCDT}

\end{document}